\documentstyle[11pt,emulateapj5]{aastex}
\def\gax{\mathrel{\raise.3ex\hbox{$>$}\mkern-14mu\lower0.6ex\hbox{$\sim$}}}
\def\lax{\mathrel{\raise.3ex\hbox{$<$}\mkern-14mu\lower0.6ex\hbox{$\sim$}}}
\def\gtorder{\mathrel{\raise.3ex\hbox{$>$}\mkern-14mu
             \lower0.6ex\hbox{$\sim$}}}
\def\ltorder{\mathrel{\raise.3ex\hbox{$<$}\mkern-14mu
             \lower0.6ex\hbox{$\sim$}}}

\begin{document}

\title{A New Mass Degeneracy in Gravitational Lenses:
Is there a crisis between a large $H_0$ and 
the time delay by a large lens halo?}
\author{HongSheng Zhao\altaffilmark{1,2} and Bo Qin\altaffilmark{1,2}}
\altaffiltext{1}{National Astronomical Observatories, Chinese Academy of 
Sciences, Beijing 100012, PRC}
\altaffiltext{2}{Institute of Astronomy, Madingley Road, Cambridge, CB3 0HA, U.K.}
\submitted{Submitted to ApJ, 18 Aug 2002}
\begin{abstract}
Isothermal models and other simple smooth models of dark matter halos
of gravitational lenses often predict a dimensionless time delay
$H_0\Delta t$ much too small to be comfortable with the observed time
delays $\Delta t$ and the widely accepted $H_0$ value ($\sim 70$ km/s/Mpc).
This conflict or crisis of the CDM has been highlighted by
several recent papers of Kochanek, who claims that 
the standard value of $H_0$ favors a strangely small halo as compact as the 
stellar light distribution with an overall nearly Keplerian rotation curve.  
In an earlier paper we argue that this is not necessarily the case, at least
in a perfectly symmetrical Einstein cross system (Paper I, astro-ph/0209191).
Here we introduce a {\it new mass degeneracy}
of lens systems to give a realistic counter example to Kochanek's claims.  We
fit the time delay and image positions in 
the quadruple image system PG1115+080.  
Equally good fits are found between lens models with flat vs. Keplerian 
rotation curves.  Time delays in both types of models can be fit with
the standard value of $H_0$.
We demonstrate that it may still be problematic
to constrain the size of lens dark halos
even if the data image positions are accurately given and the 
cosmology is precisely specified.
\end{abstract}


\section{Introduction}

Gravitational lensing provides a powerful tool to constrain the dark matter
halos of galaxies.  One of the promises of gravitational lenses is to
constrain the Hubble constant.  However, this has been hampered to some
extent by the intrinsic degeneracies in models of the dark matter
potential of the lens (Williams \& Saha 2000; Saha 2000; 
Saha \& Williams 2001; Zhao \& Pronk 2001).
Now that the value of $H_0$ is fairly well constrained by independent
methods, e.g., $H_0=72\pm 8$~km/s/Mpc from the HST key project 
(Freedman et al. 2001), and the cosmological model has been determined
at more and more precision, 
it is interesting to ask whether
we can reverse the game and set more stringent constraint on the dark
matter potential.  To this end, we would like to understand whether
the Hubble constant and the lensed images could uniquely specify the dark
matter content, or whether there are very different lens models with
identical $H_0$ value.

It is well-known that isothermal models and other simple smooth models
of dark matter halos of gravitational lenses often predict a
dimensionless time delay $H_0\Delta t$ much too small to be
comfortable with the observed time delays $\Delta t$ and the widely
accepted $H_0$ value ($\sim 70$ km/s/Mpc).  Models with isothermal dark halos
tend to yield an $H_0$ around 50 km/s/Mpc.  This conflict has been
highlighted by several recent papers (Kochanek 2002a,b,c).  Kochanek
(2002a) found that it is difficult to reconcile the time delays
measured for five simple and well-observed gravitational lenses with
$H_0 \sim 70$ km/s/Mpc unless the lens galaxy has a nearly Keplerian rotation
curve with the halo following the stellar mass profile by a constant
mass-to-light ($M/L$) ratio.  If the lenses had a more plausible flat
rotation curve (isothermal mass distributions) he found
$H_0=48_{-4}^{+7}$~km/s/Mpc, which is grossly inconsistent with the
HST Key Project.  Kochanek (2002c) argued that more realistic models
with a CDM halo plus adiabatically cooled baryons behave like
isothermal models.  They produced a still too low $H_0$ unless one
adopts a problematically high baryon fraction $\Omega_b/\Omega_m >
0.2$ of the universe, and require all these baryons to cool.  His
conclusion was that either $H_0 \sim 70$ km/s/Mpc is too high or any lens mass
models for the observed time delay systems must follow a compact
distribution, nearly like that of the stellar light, hence has very
little extended dark matter halo.  This argued for the first time a
{\it new problem} for dark matter halos, and a particularly serious
problem for current CDM paradigm of galaxy formation.

Here we discuss the effect of a {\it new degeneracy} in strong lensing
models in resolving this {\it new problem}.  It is shown that the
observed time delays and image positions cannot uniquely determine the
extent of the lens mass distribution.  In particular, a system with a
very extended dark matter distribution could minic a system without
any dark matter as far as strong lensing data are concerned.  Here we
give an analytical explicit illustration of the degree of degeneracy
in lens models.

\section{Models}

Consider fitting a general quadruple image system with the images at 
cyclindrical radius $(R_i,\theta_i)$ for $i=1,2,3,4$ 
from the lens center; we shall call the radius to the outermost image
the Einstein radius $R_E$.  These images lie at extrema points of the
time delay surface.
All lensing properties can be derived from the time delay surfaces.
As a minimal model, let's consider a spherical lens potential 
$\phi(R)=\phi_*(R)+\phi_h(R)$ for the stars and the halo 
plus an external shear potential $\phi_e(X,Y)$,
the time delay contours are determined by 
a dimensionless time delay $\tau(X,Y)$ given by
\begin{eqnarray}
\tau & = & {t \over 100h_0^{-1} \zeta(\Omega,z_l,z_s) {\rm day}\, {\rm arcsec}^{-2}}  \\
&=& {(X-X_s)^2 +(Y-Y_s)^2\over 2}- \phi_e - \phi_*-\phi_h,
\end{eqnarray}
where $R=\sqrt{X^2+Y^2}$ is the cylindrical radius, all in units of arcsec,
and $\zeta(\Omega,z_l,z_s) \sim 1$ is a constant containing 
all the dependence on the cosmology $\Omega$ and
lens and source redshifts $z_l,z_s$.

\begin{figure}
\plotone{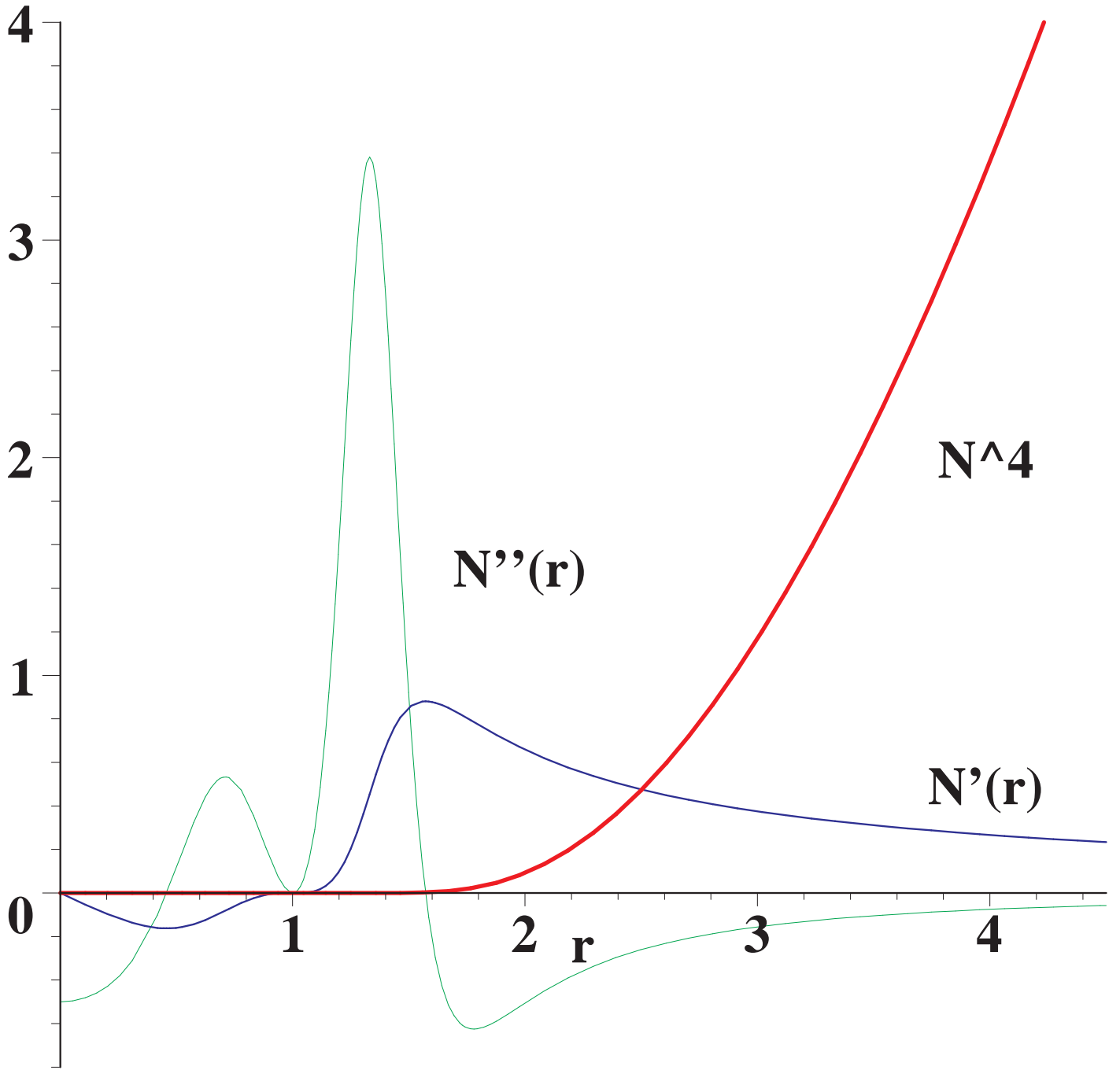}
\caption{The function $N^4(r)$ and the first and second derivatives
$N'(r)$ and $N''(r)$.
The images are at the rescaled radius $r=1$.
}
\label{N}
\end{figure}

A minimal model for the external shear is
\begin{equation}
\phi_e(X,Y) = {\gamma_1 (X^2-Y^2) + 2 \gamma_2 X Y \over 2}.
\end{equation}
A minimal model for the stars and halo would be a spherical lensing potential
\begin{eqnarray}
\phi_*(R) &=& {(1-\nu) m_0 \over \alpha}\ln \left(1+{R^\alpha \over R_e^\alpha}\right) \\
\phi_h(R) &=& {\nu m_0 \over \alpha} 
\ln \left(1+{R^\alpha \over R_e^\alpha}\right) + \phi_\nu(R)
\end{eqnarray}
where  $(1-\nu)m_0$ is 
the total stellar mass enclosed, $R_e$ is the half-mass radius,
and $2-\alpha$ specifies the cuspiness of the stellar distribution.
Here the halo mass distribution follows the mass of the stars with 
an adjustable parameter $0 \le \nu<1$,
apart from a to-be-determined halo component $\phi_\nu(R)$.  The latter
is given by
\begin{equation}
\phi_\nu(R) = m_0 \nu 
N\left({R \over R_1}\right)N\left({R \over R_2}\right)
N\left({R \over R_3}\right)N\left({R \over R_4}\right),
\end{equation}
where $R_i$ are respectively the radii of the images $i=1,2,3,4$, 
and the function $N(r)$ is given by
\begin{eqnarray}
N\left(r\right) &\equiv &
{1 \over 8} \ln \left[ 1+ \left(r^2-1\right)^4 \right]\\
&\sim & 0.0866\left(1-2r^2\right) \qquad \mbox{~if $r \ll 1$,}\\
&\sim & O[(r-1)^3] \qquad \mbox{~if $r \rightarrow 1$,}\\
&\sim & \ln r \qquad \mbox{~if $r \gg 1$}.
\end{eqnarray}
Figure~\ref{N} illustrates the function $N^4(r)$ and the first and second
derivatives of $N(r)$.  

The potential $\phi_\nu$ is designed to have the following property:
It dominates the stellar potential $\phi_*(R)$ at radii much larger
than  the Einstein radius $R_E$ with
\begin{equation}
\phi_\nu \sim m_0 \nu \ln^4(R/R_E), \qquad \phi_*(R) \sim (1-\nu)m_0 \ln(R/R_E).
\end{equation}  
But at small radius $\phi_nu$ is small and has virtually no effect on lensing. 
In particular, 
it keeps {the convergence $\kappa$, shear $\gamma$, the components of the
amplification matrix $\mu$ and the time delay between images unchanged}
at the radii of the four images $R=R_i$.  In fact
it has a vanishing contribution to 
{\it the arrival time surface near the images} up to the order
$O(R-R_i)^3$.  This can be verified by Taylor expansion.  
It is easy to show (also cf. Fig.~\ref{N}) that the factor 
$N(r)$ and its first and second radial derivatives obey
\begin{equation}
\left.{N''}\right|_{R=R_i} =
\left.{N}\right|_{R=R_i} = 
\left.{N}\right|_{R=R_i} = 0.
\end{equation}
Becaue $\phi_\nu(R)$ is a multiplications of the four $N(r)$,
all of the zeroth, first and second derivatives vanish at the four 
radii $R_i$.
And because $\phi_\nu(R)$ is azimuthally symmetric, all its 
azimuthal derivatives with respect to $\theta$ also vanish.

The time delay surface
\begin{eqnarray}
\tau(X,Y) &=& {(X-X_s)^2 +(Y-Y_s)^2\over 2} \nonumber\\
&-&{\gamma_1 (X^2-Y^2) + 2 \gamma_2 X Y \over 2} \nonumber\\
&-&{m_0 \over \alpha} \ln \left(1+{R^\alpha \over R_e^\alpha}\right) \nonumber\\
&-&m_0 \nu \Pi_{i=1}^{4} N\left({R \over R_i}\right).
\end{eqnarray}
Hence when we vary the parameter $\nu$,
the time arrival surface $\tau(X,Y)$ yields the same extrema, or images.
This means models with different $\nu$ (or halo) will have goodness of fit to 
all strong lensing data (the time delay, the image positions and
even the amplifications and parity of the images).  It only alters the mass
distribution of the lens, e.g., the total mass of the lens and the
spatial extent of the lens.  Hence it creates a degeneracy in lens
modeling, making it
problematic to draw unique conclusions on lens halo mass from image modeling.
We note, however, by construction this degeneracy 
applies only to point images or point sources.  

The surface density and mass of the model can be computed as
\begin{equation}
\kappa(R) = 1-{1 \over 2} \nabla^2 \tau,\qquad 
M(R) =R {d \phi \over d R},
\end{equation}
in particular, the density and the mass for the stars only are given by
\begin{equation}
\kappa_*(R) =  {\alpha  (1-\nu) m_0
R_e^\alpha \over 2 R^{2-\alpha} (R^\alpha+R_e^\alpha)^2}
\end{equation}
and
\begin{equation}
M_*(R) = {(1-\nu)m_0R^\alpha \over R^\alpha+R_e^\alpha}.
\end{equation}
As we can see the mass in stars has a half-mass radius $R=R_e$,
and converges to a finite mass $m_0$ at infinity.
The density has an inner cusp $2-\alpha$, and drops steeply with radius, 
by a factor of 25/4 from $R=R_e$ to $R=2R_e$ for cored models ($\alpha=2$)
or by a factor of 4 for isothermal cusped models ($\alpha=0$).

The halo component $\phi_\nu(R)$ contributes
to $\kappa(0)$ by a negligible 
amount $\pm 0.001 \nu$ for $0 \le R \le 1.3R_E$, but 
contributes to $\kappa(20R_E>R>2R_E)$ by an amount of the order $\nu R_E/R$.
This flattens the rotation curve at large radii.  
A flat rotation curve corresponds to a constant deflection strength, 
\begin{equation}
u^2 \equiv {d \phi \over d R} \equiv {M \over R}, 
\end{equation}
which is effectively the rotation curve squared.  
At large radius we have
\begin{equation}
M \rightarrow m_0+3\nu m_0 \ln^3(R/R_E).
\end{equation}
In comparsion to the case with only the stars, we have
a nearly Keplerian rotation curve beyond $R=2R_e$.

\section{Results}

PG1115+080 is a quadruple system with a nearly axisymmetric stellar
lens at $z_l=0.31$, and the quasar source is at $z_s=1.72$.  
We use the photometric data of Impey et al. (\cite{Impey98}) for the
images A1, A2, B, C, and the time delay $t_{BC}=25$ days 
between the nearest B image and the furthest C image (Schechter et al. 1997).  
There is an infared Einstein ring of radius $R_E \sim 1.4\arcsec$.
The stellar lens is well-approximated by a de Vaucouleurs profile 
with a half-light radii of $R_e=0.55\arcsec=0.4R_E$.  Since the surface density
is nearly cored, we set the cusp slope $2-\alpha=0$.  
We adopt a $\Lambda$CDM cosmology with the time delay constant
for at the redshifts $z_l=0.31,z_s=1.72$ of PG1115+080 given by
\begin{equation}
100h_0^{-1}\zeta(\Omega=1-\Lambda) \sim (30-32)h_0^{-1}{\rm day}\,{\rm arcsec}^{-2}.
\end{equation}

\begin{figure}
\plottwo{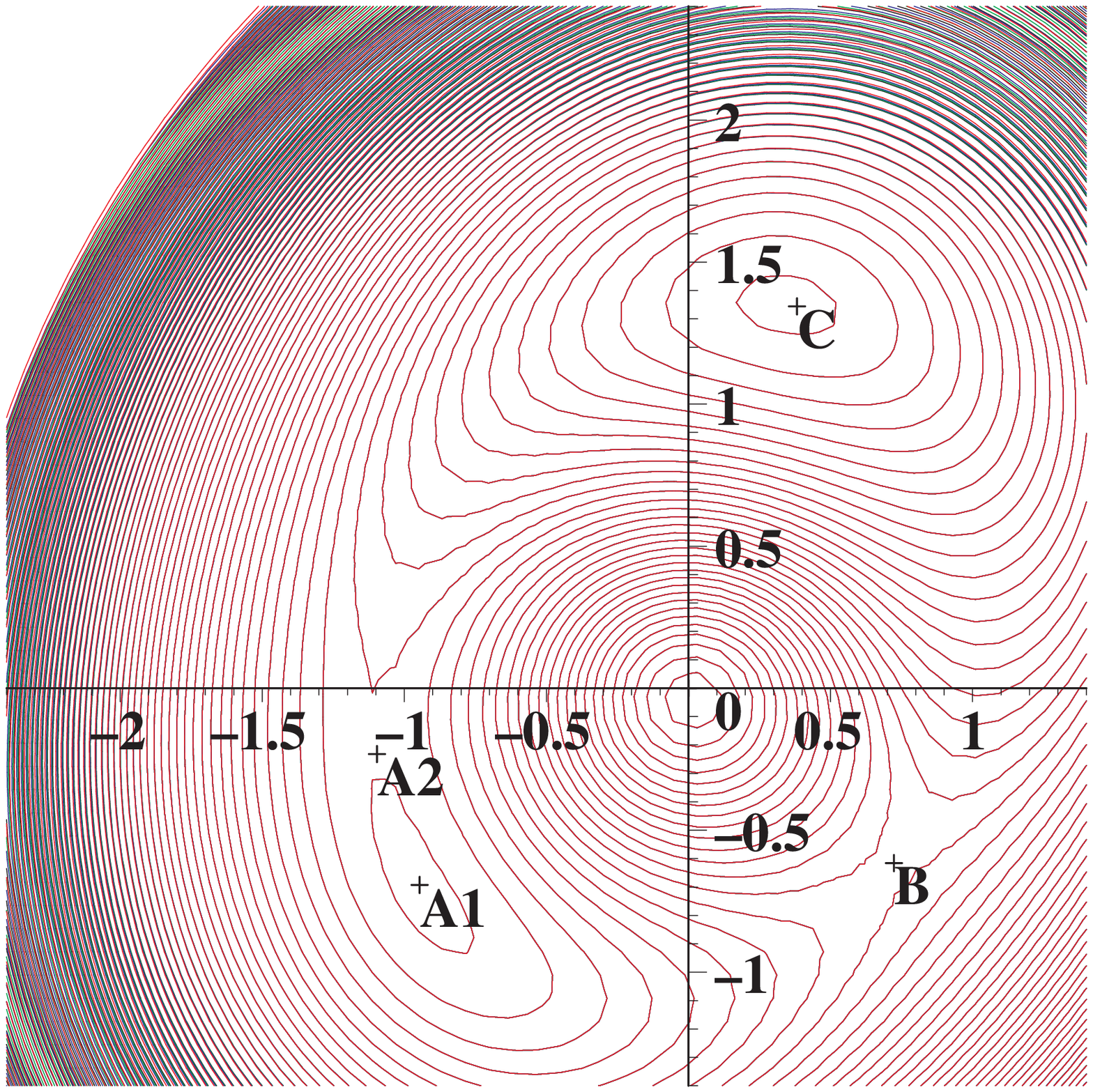}{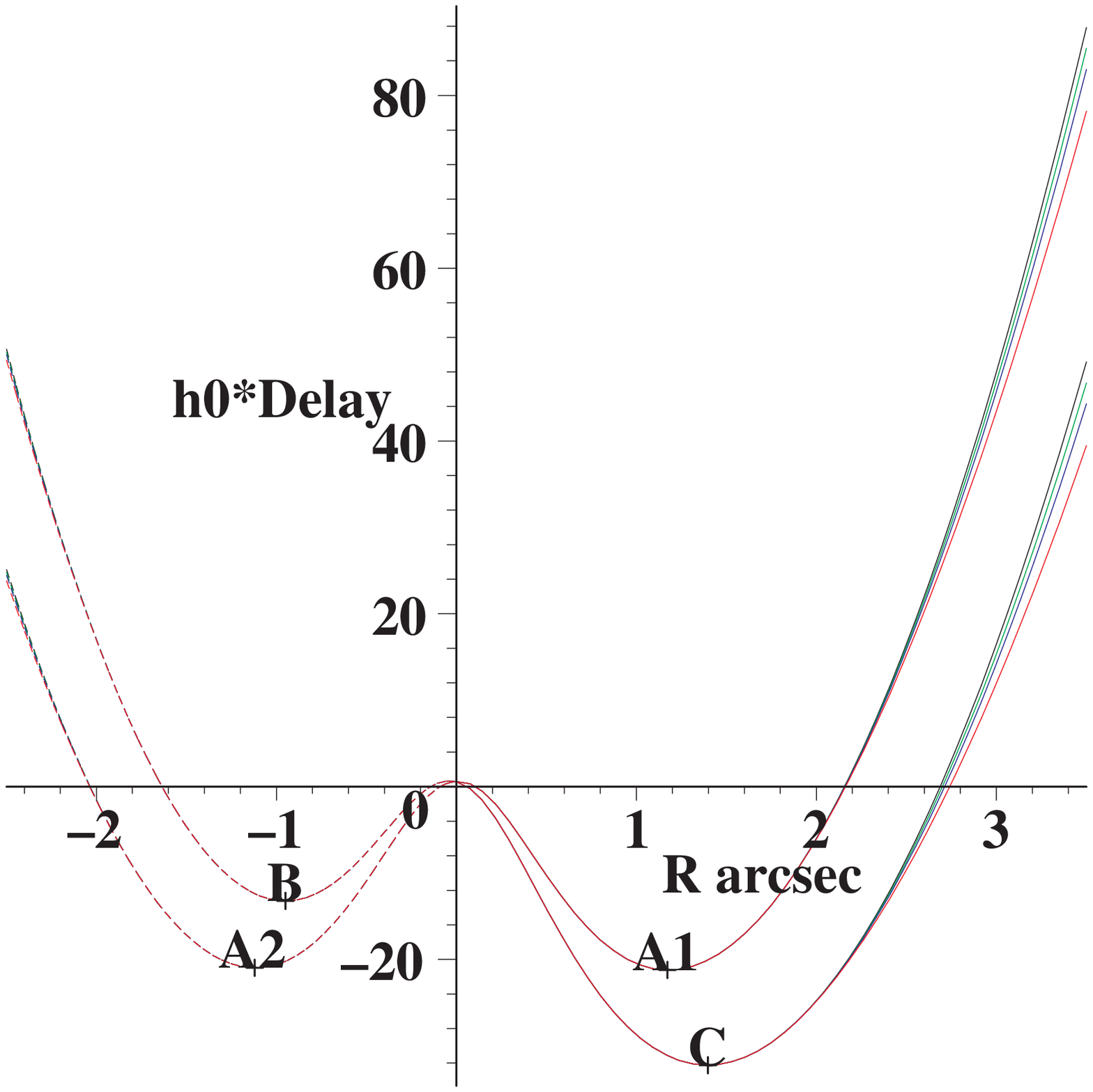}
\caption{
Panel (a) shows 
time delay contours at intervals of 1 day for the four lens models.
All models reproduce the same image positions and time delay with $h_0=0.72$.
The difference of the models starts to be visible 
(the contours become thicker) beyond 3 arcsecs from the center. 
Panel (b) shows cuts of the time delay surfaces of the four lens models
along the radial direction from the lens to the time delay minima
(marked by ``C'' and ``A1''on solid curves) 
and from the lens to the saddle images (marked by ``B'' and ``A2'' 
on dashed curves). From bottom to top lines are models
with decreasing amount of halos $\nu=0.16,0.08,0.04,0$
with color coded as red, blue, green and black.  
The measured delay $t_{BC}=25$ days
implies $h_0=0.72$ for all these models. 
The time delay ratio $t_{AC}/t_{BA}=1.13\pm 0.18$ of Barkana (1997) is 
also reproduced.
}\label{ctt}
\end{figure}

Let's try to fit PG1115 with a halo-free model with $\nu=0$.  Totally we
have five free fitting parameters, with $(\gamma_1,\gamma_2,m_0)$ for
the lens model, and $(X_s,Y_s)$ being a pair of coordinates for the source.
Since the four images provide in general at least eight constraints,
we do not expect a perfect fit.  Nonetheless a reasonable fit can be
found for this star-only model, as shown by the time delay contours in
Figure~\ref{ctt}a and the cuts in Figure~\ref{ctt}b.  The model yields 
a time delay 
\begin{equation}
\tau_{BC} \sim 0.56~{\rm arcsec}^2,
\end{equation}
between images B and C.  Compared with measured delay 
$t_{BC}=25{~\rm day}$, we find
\begin{equation}
h_0={100\zeta \tau_{BC} \over t_{BC}} \sim 
{30 \tau_{BC} \over t_{BC}} \sim 0.72
\end{equation} 
for this star-only model.  The high value of $h_0$ is 
related to the very small convergence at the Einstein radius $R_E$.

\begin{figure}
\plotone{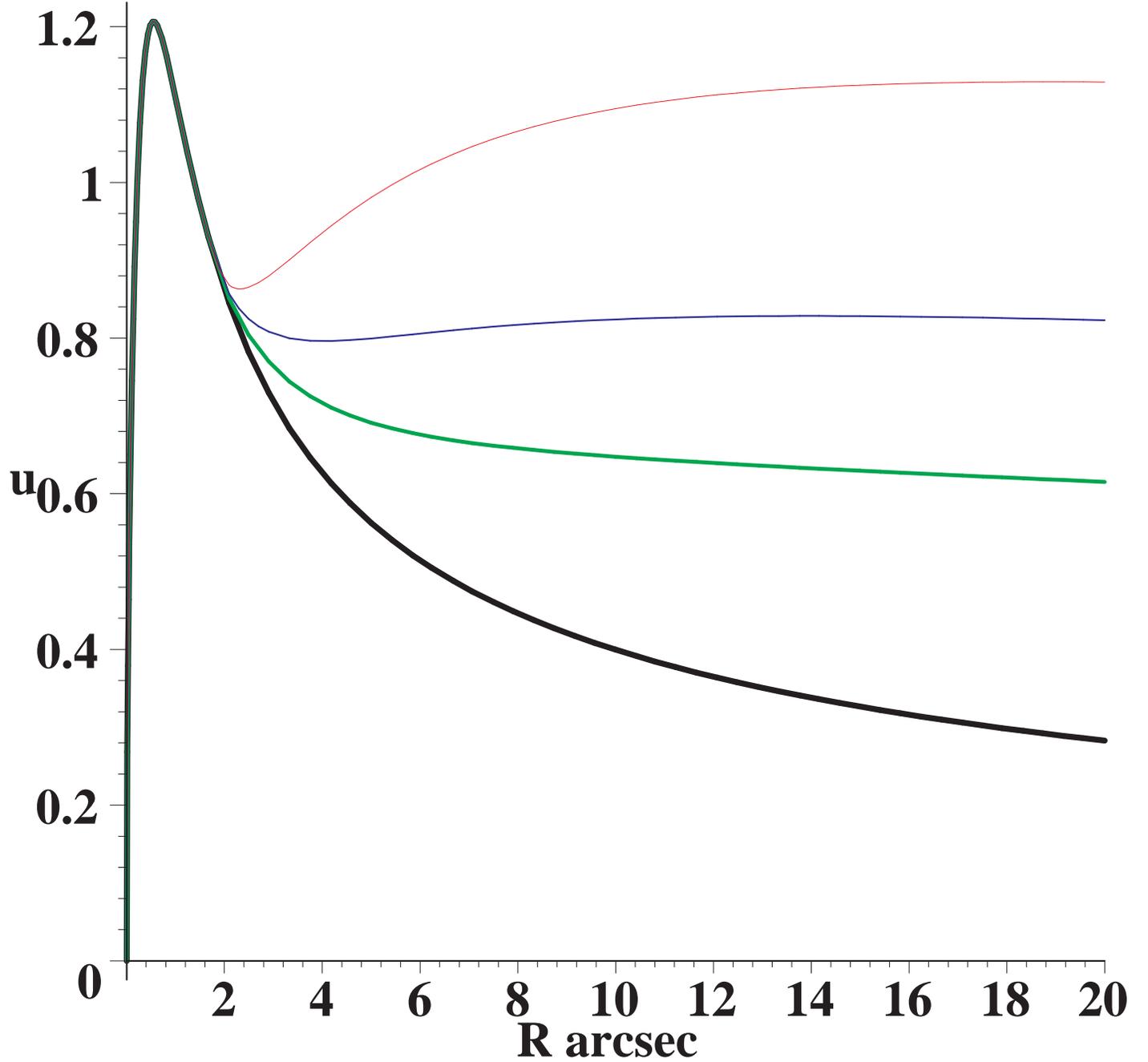}
\caption{The deflection strength of the lens 
$u^2={M(R) \over R}={d \phi \over dR}$ as a function of the distance
from the lens; this is effectively a rotation curve of the lens.
From top right to down right, from thinner to thicker lines are models
with decreasing amount of halos $\nu=0.16,0.08,0.04,0$
with color coded as red, blue, green and black.
Note that both Keplerian and flat rotation curve models are allowed by 
the data.}
\label{v}
\end{figure}

Now consider adding the component $\phi_\nu$, i.e., we increase the value
$\nu$ from zero to $\nu=0.04,0.08,0.16$.  We see no detectable
differences in the time delay (cf. Fig.~\ref{ctt}a and Fig.~\ref{ctt}b),
hence we still have $h_0=0.72$.  All models predict amplification
ratios among the four images to be
\begin{equation}
A_1:A_2:B:C=3:4:0.7:1,
\end{equation}
independent of $\nu$.  These are in good agreement with 
the observed infared flux ratios
\begin{equation}
A_1:A_2:B:C=3.84:2.67:0.64:1,
\end{equation}
apart from the well-known problematic flux ratio between the close
pair $A_1$ and $A_2$, which has been attributed to either microlensing
or lensing by substructures (e.g., Impey et al. 1998; 
Barkana 1997; Metcalf \& Zhao 2002).
Consistent with Kochanek (2002b), 
our model density is small near $R_E$ with a convergence
\begin{equation}
\kappa(R_E) \sim 0.1 + 0\times \nu,
\end{equation}
independent of $\nu$; 
by construction $\phi_\nu$ has a vanishing contribution
to the convergence at the images.  But by raising the parameter $\nu$ 
the models develop a very massive halo at radii $\gg R_E$ 
with a nearly flat rotation curve (cf. Fig.~\ref{v}),
very different from the model with $\nu=0$.  This illustrates an explicit 
counter example to Kochanek's claim of a conflict between 
lens models with an extended dark halo and $\Lambda$CDM cosmology with 
$H_0\sim 70$.  We do not find such a conflict.  Instead
we find that the new mass degeneracy prevents us from 
drawing a robust conclusion 
about the dark halo on the basis of the image positions and time delays alone.

\section{Summary}

As we can see, it is possible to construct many very different models
with positive, smooth and monotonic surface densities to fit the image
positions.  There are also no extra images.  These models
also fit the same time delay and time delay ratios using 
a Hubble constant and cosmology consistent with $\Lambda$CDM cosmology.
Hence the models are truely indistinguishable for lensing data.
They fit the flux ratios equally well, and 
produce nearly indistinguisable Einstein rings, which is the region of 
minimal gradient of the time delay surface.  The models have identical
light profiles, undistinguishable by data.  

Among the acceptable models to PG1115+080, there are models with a Keplerian
rotation curve and models with a nearly flat rotation curve.  So lensing
data plus $H_0$ cannot uniquely specify the mass-to-light ratio of this
system.

We conclude that strong lensing data may not uniquely determine the
Hubble constant, even if we fix the cosmology, the lens and source
redshifts, and the time delays and amplification ratios of the four
images.  There are at least important degeneracies in inverting the
data of a perfect Einstein cross to the lens models and the Hubble
constant.  The relation between the value of $H_0$ and
the size of the halo is not straightforward: 
a high $H_0$ does not necessarily mean no dark halo, and models
with a flat rotation curve do not always yield a small $H_0$.  We
also comment that it would be difficult to determine the cosmology from
strong lensing data alone because the non-uniqueness in the lens
models implies that the combined parameter $h_0^{-1}\zeta(\Omega,z_l,z_s)$ is
poorly constrained by the lensing data, even if $h_0$ and the
redshifts $z_l,z_s$ are given.

This work was supported by the National Science Foundation
of China under Grant No. 10003002 and a PPARC rolling
grant to Cambridge.  HSZ and BQ thank the Chinese Academy of Sciences
and the Royal Society respectively for a visiting fellowship, and the
host institutes for local hospitalities during their visits.



\begin{deluxetable}{rcccccccr} 
\tablecolumns{9} 
\tablewidth{0pc} 
\tablecaption{Plausible lens parameters to fit the images of 
a perfect Einstein cross and time delays with $H_0=72$ km/s/Mpc
}
\tablehead{ 
\colhead{Size} 
&\colhead{Mass}
&\colhead{Shear} 
& \colhead{Source} 
& \colhead{Convergence}
& \colhead{}
& \colhead{}
& \colhead{}
\\
\colhead{$R_e/R_E$} 
&\colhead{$m_0$}
&\colhead{$(\gamma_1,\gamma_2)$} 
& \colhead{$(X_s,Y_s)$} 
& \colhead{$\kappa(0)$}
& \colhead{$\kappa(R_E)$}
& \colhead{$\kappa(10R_E)$}
& \colhead{$\kappa(100R_E)$}
}
\startdata
0.55/1.4 & 1.6 & (-0.11,0.14) & (-0.05,0.18) & 
$5.3-1.6\times 10^{-3}\nu$ &
$10^{-1}$ & 
$10^{-4}+0.3\nu$ & 
$10^{-9}+0.01\nu$ 
\enddata 
\end{deluxetable} 

\end{document}